\documentclass[aps,prb,twocolumn,superscriptaddress,showpacs]{revtex4-1}

\usepackage{graphicx}
\usepackage{amsfonts}
\usepackage{amsmath}
\usepackage{amssymb}
\usepackage{bm}
\usepackage{color}

\begin{document}

\title{Collective magnetic and plasma excitations in Josephson $\psi$ junctions}
\author{S. V. Mironov}
\affiliation{Institute for Physics of Microstructures, Russian Academy of Sciences, 7 Academicheskaya Str., Nizhniy Novgorod 603087, Russia}
\author{A. I. Buzdin}
\affiliation{University Bordeaux, LOMA UMR-CNRS 5798, F-33405 Talence Cedex, France}
\affiliation{World-Class Research Center ``Digital biodesign and personalized healthcare'', Sechenov First Moscow State Medical University, Moscow 119991, Russia}

\begin{abstract}
We show that Josephson $\psi$ junctions with the half-metallic (HM) weak link coupled to the superconducting (S) electrodes through the ferromagnetic (F) layers host collective excitations of magnetic moment and the Josephson phase. This results in the shift of the ferromagnetic resonance frequency, anomalies in the current-voltage characteristics and appearance of additional magnetic anisotropy in the F layers. In contrast to the previously studied S/F/S junctions, the coupling between magnetic and plasma modes emerges even in the long wavelength limit. Such coupling is shown to enable the controllable magnetization reversal in the F layer governed by the d.c. current pulse which provides the effective mechanism for the magnetic moment manipulation in the devices of superconducting spintronics. 
\end{abstract}

\maketitle

\section{Introduction}\label{Sec_Intro}

Josephson junctions with ferromagnetic (F) interlayers between the superconducting (S) electrodes are known to support the variety of exotic quantum phenomena.\cite{Golubov, Buzdin_RMP}. The conversion of spin-singlet Cooper pairs into the triplet ones at S/F interfaces results in the anomalous local increase in the electronic density of states at the Fermi level,\cite{Buzdin_DOS} appearance of the long-range Josephson currents,\cite{BVE_LRTC, Kadigrobov, Robinson_LRTC, Khaire} formation of $\pi$ junctions\cite{Buzdin_pi, Ryazanov_1, Ryazanov_2} etc. Even more unusual phenomena arise in the S/F/S multilayered systems with broken inversion symmetry resulting in the strong spin-orbit coupling (SOC). The most remarkable feature of such sandwiches is the so-called anomalous Josephson effect, i.e. emergence of the nonzero phase $\varphi_0\not= \pi$ between the superconducting electrodes in the ground state so that near the superconducting critical temperature the relation between current $I$ and phase $\varphi$ takes the form $I=I_c\sin(\varphi-\varphi_0)$ (where $I_c$ is the critical current).\cite{Buzdin_Phi, Reynoso, Zazunov, Mironov_Phi, Kouwenhoven_Phi} Being integrated into the superconducting loops such $\varphi_0$ junctions play the role of phase batteries producing the spontaneous currents,\cite{Ustinov, Bauer, Buzdin_2005, Feofanov, Ortlepp} which is expected to bring new functionality to the devices of the rapid single-flux quantum logics.\cite{Likharev_RSFQ} 

During the past decade both conventional S/F/S and $\varphi_0$ junctions acquired much attention also because of their unusual dynamic properties. Specifically, in multilayered S/F/S junctions the spin waves in the F layer becomes coupled to the dynamics of the Josephson phase difference $\varphi$ (plasma-like waves) provided the propagation vector along the junction plane is nonzero.\cite{Volkov_1} Experimentally, the hybridized waves are expected to reveal themselves mainly through substantial changes in the number, position and shape of Fiske and Shapiro steps on the I-V characteristics of S/F/S junctions.\cite{Volkov_1, Chudnovsky_1, Volkov_2, Maekawa_1, Belzig, Maekawa_2, Maekawa_3, Ebrahimi, Nashaat, Sherbini, Shukrinov_2} Another remarkable manifestation of such coupling between plasma-like and spin waves is the possibility to stimulate the magnetization reversal in the F layer by applying the current through the junction.\cite{Chudnovsky_1, Linder, Hoffman, Chudnovsky_2, Bobkova} This may provide the effective mechanism of the magnetic moment control in the devices of superconducting spintronics.\cite{Linder_rev, Eschrig_rev} Similar hybridization of  Josephson phase oscillations and precession of magnetic moment is predicted for the $\varphi_0$ junctions.\cite{Konschelle, Chudnovsky_3, Chudnovsky_4, Shukrinov_1, Shukrinov_3, Shukrinov_4, Shukrinov_5, Guarcello} The ground state phase $\varphi_0$ is determined by the angle $\theta$ between the magnetic moment ${\bf M}$ orientation inside the ferromagnet and the unit vector ${\bf n}$ along the direction of the broken inversion symmetry so that $\varphi_0\propto\sin\theta$. As a result, the Josephson energy $E=(\Phi_0I_c/2\pi c)\left[1-\cos(\varphi-\varphi_0)\right]$ becomes dependent on the magnetic moment orientation which results not only in the influence of the magnetic order on the superconducting current but also in the back-action of the Josephson current on the magnetization direction. Consequently, outside the equilibrium magnetic and superconducting excitations in the $\varphi_0$ junction becomes coupled to each other giving rise to the collective oscillations. Similarly to the S/F/S junctions, the collective plasma-like and spin excitations provide the possibility to realize the ultra-fast reversal of magnetization direction controlled by the short current pulses which is promising for the design of memory cells.  

The further advance in the field of spontaneous Josephson effect is associated with the implementation of fully spin-polarized ferromagnets which are often called half-metals (HM).\cite{Pickett_PT, Coey_JAP} Although the singlet Cooper pairs consisting of two electrons with opposite spins cannot penetrate half-metal directly, the experiments on the S/HM/S systems demonstrate the existence of the Josephson transport through the HM layer.\cite{Keizer, Anwar} This unusual observation is attributed to the spin-active interfaces of the half-metal which transform the spin structures of Cooper pairs converting them from spin-singlet state to the triplet one.\cite{Eschrig_PRL_2003, Eschrig_PhysToday} The controllable singlet-triplet conversion can be achieved in the complex multilayered S/F/HM/F/S structures with non-coplanar orientation of magnetic moments in the three ferromagnetic layers (see, e.g., Refs.~[\onlinecite{Zheng, Feng, Linder_HM, Enoksen, Asano_HM, Beri}]). The theoretical calculations within different approaches predict the anomalous Josephson effect for the S/F/HM/F/S junctions with the current-phase relation of the form $I(\varphi)=I_c\sin(\varphi-\psi-\pi)$ where the critical current $I_c>0$, and $I_c\propto\left|\sin\vartheta_1\sin\vartheta_2\right|$ ($\vartheta_1$ and $\vartheta_2$ are the angles between the magnetic moments in the F$_1$ and F$_2$ layers and the $z$ axis). Remarkably, the spontaneous phase $\psi$ is equal to the angle between the projections of magnetic moments in the two F layers to the plane perpendicular to the spin quantization axis of half-metal and does not depend on all other system parameters.\cite{Nazarov, Eschrig_2009, Eschrig2015} Note that for the parallel orientation of the magnetic moments projections the junction should be in the $\pi$ state.\cite{Nazarov, Asano_2007, Liu_2010} This additional $\pi$ shift can be accounted with the minus sign in the current-phase relation so that it takes the form 
\begin{equation}\label{CPR}
I(\varphi)=-I_c\sin(\varphi-\psi).
\end{equation}
The most striking feature of the S/F/HM/F/S $\psi$ junctions which contrasts to the properties of usual $\varphi_0$ junctions is the Josephson phase accumulation accompanying the magnetic moment rotation in one of the F layers. Recently, this Berry phase effect was shown to allow the controllable pumping of magnetic flux into the superconducting loop containing $\psi$ junction without application any out-of-plane external magnetic field.\cite{Mironov_2020} 
  
In this paper we show that the peculiar coupling between magnetic moments and Josephson phase in the $\psi$ junctions gives rise to the collective magnetic and plasma excitations which substantially differ from the ones in the conventional S/F/S systems and $\varphi_0$ junctions. Specifically,  in contrast to the S/F/S structures $\psi$ junctions enable the coupling between magnetic and plasma oscillation even for the zero wave-vector along the junction, which results in the measurable shifts in the ferromagnetic resonance frequency as well as in the anomalies on the current-voltage characteristics. Moreover, we show that in the rf SQUID geometry when the S/F/HM/F/S $\psi$ junction is integrated into the superconducting loop the coupling between the Josephson phase and magnetic degrees of freedom effectively renormalized the anisotropy in one of two F layers producing an additional easy axis direction. Finally, we demonstrate the controllable magnetization reversal in one of the F layers under the effect of the external d.c. current pulse flowing through the junction, which provides an effective tool for the magnetic moment manipulation in the devices of superconducting spintronics. 

The paper is organized as follows. In Sec.~\ref{Sec_Excitations} we calculate the resonance frequencies of the collective magnetic and plasma excitations in S/F/HM/F/S junction and compare the results with the ones previously obtained for the S/F/S systems. In Sec.~\ref{Sec_Anisotropy} we analyzed the additional induced magnetic anisotropy in the $\psi$ junction integrated into the superconducting loop. In Sec.~\ref{Sec_magn_control} we demonstrate the magnetic moment reversal in the S/F/HM/F/S junction under the effect of the d.c. current. Finally, in Sec.~\ref{Sec_conclusion} we summarize our results.

\section{Collective excitations in an isolated Josephson junction}\label{Sec_Excitations}

\begin{figure}[t!]
\includegraphics[width=0.48\textwidth]{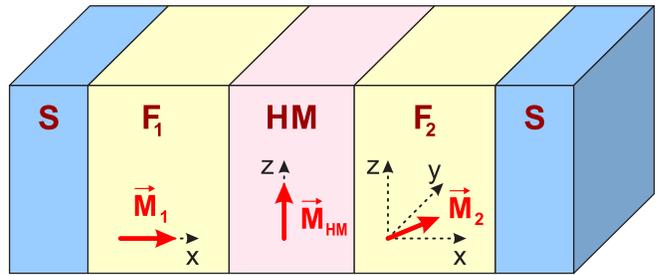}
\caption{Sketch of the Josephson $\psi$ junction based on the multilayered S/F/HM/F/S structure.}\label{Fig1}
\end{figure}

We start from the analysis of the collective magnetic and superconducting excitations in the isolated S/F$_1$/HM/F$_2$/S junction where the spin quantization axis in the half-metal is directed along the $z$  axis, the magnetic moment in the F$_1$ layer is fixed along the $x$ axis while the magnetization ${\bf M}$ in the F$_2$ layer can change its direction (see Fig.~\ref{Fig1}). For simplicity we will restrict ourselves to the case when the temperature is well below the Curie temperature so that the absolute value of the magnetization $M$ has reached the saturation value $M_0$ and is almost constant. 

To describe the dynamics of the Josephson phase we use the resistively shunted junction (RSJ) model introducing the normal state resistance $R_N$ and the capacity $C$ of the junction:\cite{} 
\begin{equation}\label{RSJ}
\frac{\partial^2\varphi}{\partial t^2}+\frac{\omega_p}{Q}\frac{\partial\varphi}{\partial t}-\omega_p^2\sin\left(\varphi-\psi\right)=0,
\end{equation}
where $\omega_p=\sqrt{2\pi I_c c/(\Phi_0 C)}$ is the plasma frequency, $Q=\omega_p R_N C$ is the quality factor.

The magnetization dynamics can be described by the Landau-Lifshitz-Gilbert equation:
\begin{equation}\label{LLG}
\frac{\partial{\bf M}}{\partial t}=\gamma\left[{\bf H}_{eff}\times{\bf M}\right]+\frac{\alpha}{M_0}\left[{\bf M}\times\frac{\partial{\bf M}}{\partial t}\right],
\end{equation} 
where $\gamma$ is the electron gyromagnetic ratio, $\alpha$ is the phenomenological damping constant, and the effective field ${\bf H}_{eff}$ is determined by the derivative of the system free energy $F$ accounting both the Josephson and magnetic contributions:
\begin{equation}\label{Heff}
{\bf H}_{eff}=-\frac{1}{V}\frac{\partial F}{\partial {\bf M}}.
\end{equation} 

The specific form of the free energy functional $F$  depends on the type of magnetic anisotropy in the F$_2$ ferromagnet. Let us consider the case when the F$_2$ later has the easy-axis anisotropy along the $x$ axis. Then the system free energy has the form
\begin{equation}\label{FE1}
F=E_J\left[1+\cos\left(\varphi-\psi\right)\right]-\frac{K_\parallel V}{2}\left(\frac{M_x}{M_0}\right)^2.
\end{equation}
Here $E_J=\Phi_0I_c/(2\pi c)$ is the Josephson energy, $K_\parallel >0$ is the magnetic anisotropy constant, $V$ is the volume of the F$_2$ layer, and $M_x$ is the projection of the magnetization ${\bf M}$ to the easy axis. 

One sees that the system ground state is degenerate twice: in equilibrium the magnetization is directed parallel or anti-parallel to the $x$-axis so that $M_x=\nu M_0$ where $\nu=\pm 1$. The corresponding values of the ground-state Josephson phase are $\varphi_0^\nu=\pi$ for $\nu=+1$ and $\varphi_0^\nu=0$ for $\nu=-1$.

In the case of strong magnetic anisotropy ($K_\parallel V\gg E_J$) the deviations of the vector ${\bf M}$ direction from the $x$ axis are small. In this case one may put $M_x\approx \nu M_0$ and the angle $\psi$ entering the current-phase relation of the junction takes the value $\psi\approx M_y/M_0$ for $\nu=1$ and $\psi\approx \pi-M_y/M_0$ for $\nu=-1$.  

Taking the derivative in (\ref{Heff}) we get the expressions for the effective field 
\begin{equation}\label{Heff_1}
{\bf H}_{eff}=\frac{K_\parallel M_x}{M_0^2}\hat{\bf x}_0-\frac{E_J}{V}\sin\left(\varphi-\psi\right)\frac{\partial \psi}{\partial M_y}\hat{\bf y}_0,
\end{equation} 
and in the vicinity of the two ground states  
\begin{equation}\label{Heff_2}
{\bf H}_{eff}^\nu=\frac{\nu K_\parallel }{M_0}\hat{\bf x}_0-\frac{ E_J}{VM_0}\sin\left(\varphi-\frac{\nu M_y}{M_0}\right)\hat{\bf y}_0.
\end{equation} 
Using the system of Eqs.~(\ref{RSJ})-(\ref{LLG}) let us first determine the resonant frequencies of the Josephson junction neglecting damping, i.e., considering the case $\alpha\to 0$ and $Q\to\infty$. It is convenient to introduce the unit vector ${\bf m}={\bf M}/M_0$ directed along the magnetization. Then considering the oscillatory process where all values $m_y$, $m_z$ and $\delta\varphi=\varphi-\varphi_0^\nu$ are proportional to $e^{i\omega t}$ and linearizing Eqs.~(\ref{RSJ})-(\ref{LLG}) for the corresponding complex amplitudes (which will be further indicated by the tilde) we obtain:
\begin{equation}\label{Eq_My}
i\omega \tilde{m}_y=-\nu\omega_{\parallel}\tilde{m}_z,
\end{equation}
\begin{equation}\label{Eq_Mz}
i\omega \tilde{m}_z=\nu\omega_{\parallel}\tilde{m}_y-\beta_\parallel \omega_{\parallel}\left(\delta\tilde{\varphi}-\nu\tilde{m}_y\right),
\end{equation}
\begin{equation}\label{Eq_phi}
-\omega^2 \delta\tilde{\varphi}+\omega_p^2\left(\delta\tilde{\varphi}-\nu\tilde{m}_y\right)=0,
\end{equation}
where $\omega_{\parallel}=\gamma K_\parallel /M_0$ is the ferromagnetic resonance frequency (we assume $\omega_{\parallel}>\omega_p$) and $\beta_\parallel=E_J/(K_\parallel V)$ is the dimensionless parameter characterizing the ratio between the Josephson and the anisotropy energies (the above assumption of strong magnetic anisotropy field requires $\beta_\parallel\ll 1$). The system (\ref{Eq_My})-(\ref{Eq_phi}) may have a non-trivial solution provided
\begin{equation}\label{w2_eq}
\omega^4-\omega^2\left[\omega_p^2+\omega_\parallel^2(1+\beta_\parallel)\right]+\omega_p^2\omega_\parallel^2=0.
\end{equation}
This equation has two real roots for arbitrary choice of the system parameters: 
\begin{equation}\label{w2_res}
\omega^2=\frac{\omega_p^2+\omega_\parallel^2(1+\beta_\parallel)\pm\sqrt{\left[\omega_p^2+\omega_\parallel^2(1+\beta_\parallel)\right]^2-4\omega_p^2\omega_\parallel^2}}{2}.
\end{equation}
In the limit $\beta_\parallel\ll (1-\omega_p^2/\omega_\parallel^2)$ the two resonance frequencies $\omega_1$ and $\omega_2$ read
\begin{equation}\label{w2_small_beta}
\omega_1\approx \omega_p\left(1-\frac{\beta_\parallel}{2}\frac{\omega_\parallel^2}{\omega_\parallel^2-\omega_p^2}\right),~~~\omega_2\approx\omega_\parallel\left(1+\frac{\beta_\parallel}{2}\frac{\omega_\parallel^2}{\omega_\parallel^2-\omega_p^2}\right).
\end{equation}

Remarkably, the hybridization of the magnetic and superconducting excitations in the S/F/HM/F/S junction substantially differs from the similar phenomenon in S/F/S junctions previously discussed in Ref.~[\onlinecite{Volkov_1}]. The key feature specific to the $\psi$ junction is the mixing of magnetic and plasma waves even at zero wave vector while in S/F/S structures such mixing arises only for the waves propagating along the junction. As a consequence, in contrast to the S/F/S junction, in the long wavelength limit the coupling between magnetic and plasma oscillation in the $\psi$ junction results in the shift of the ferromagnetic resonance frequency (see the above expression for $\omega_2$) which provides an effective tool for the experimental observation of the predicted collective excitations. Note that the lateral size of the experimentally fabricated S/F/S Josephson junctions with the composite F layer is typically small and lay in the range $0.5-20~{\rm \mu m}$ \cite{Robinson_LRTC, Khaire, Komori} which is much smaller than the typical values of Josephson length $\lambda_J\sim 100~{\rm \mu m}$. This means that in the experimentally achievable structures only uniform plasma oscillations with zero wave-vector should be generated.

Also the interaction between the oscillation of magnetic moment and Josephson phase should give rise to the anomalies in the current-voltage characteristics of $\psi$ junctions. To demonstrate this we consider the limit when the current flowing through the junction is much larger than $I_c$. In this case the voltage $U$ across the junction is almost constant so that the Josephson phase linearly depends on time: $\varphi=\omega_J t$ where $\omega_J=2eU/\hbar$. We restrict ourselves to the case of small deviations of the magnetic moment vector from its equilibrium direction assuming $m_x,~m_y\propto \beta_\parallel\ll 1$ and neglecting the contributions $\sim O(\beta_\parallel^2)$. Then in the presence of damping (for $\alpha\neq 0$) we may write the equation (\ref{LLG}) describing the dynamics of magnetic moment in the form
\begin{equation}\label{Eq_My2}
\frac{\partial m_y}{\partial t}=\frac{\omega_\parallel}{1+\alpha^2}\left[-\nu m_z-\alpha m_y -\alpha\beta_\parallel\sin(\omega_J t)\right],
\end{equation}
\begin{equation}\label{Eq_Mz2}
\frac{\partial m_z}{\partial t}=\frac{\omega_\parallel}{1+\alpha^2}\left[\nu m_y-\alpha m_z +\nu\beta_\parallel\sin(\omega_J t)\right].
\end{equation}
The solution of these equations is somewhat similar to the one previously obtained for the $\varphi_0$ junctions \cite{Konschelle}:
\begin{equation}\label{my_sol}
m_y(t)=\left(\omega_{-}-\omega_{+}\right)\sin(\omega_Jt)+\left(\alpha_{+}+\alpha_{-}\right)\cos(\omega_Jt),
\end{equation}
where 
\begin{equation}\label{const_def}
\omega_\pm=\frac{\beta_\parallel}{2\omega_\parallel}\frac{\omega_J\pm\omega_\parallel}{\Omega_\pm},~~\alpha_\pm=\frac{\alpha\beta_\parallel\omega_J}{2\omega_\parallel \Omega_\pm},
\end{equation}
\begin{equation}\label{Omega_def}
\Omega_\pm=\frac{\left(\omega_J\pm \omega_\parallel\right)^2+\left(\alpha\omega_J\right)^2}{\omega_\parallel^2}.
\end{equation}
The obtained solution for $m_y(t)$ allows us to calculate the superconducting current $I_s=-\nu I_c\sin(\omega_Jt-\nu m_y)$. Remarkably, it has a d.c. component $I_s^{dc}$ which arises due to the damping:
\begin{equation}\label{Is_dc}
I_s^{dc}=\frac{\alpha\beta_\parallel I_c\omega_J}{4\omega_\parallel }\left(\frac{1}{\Omega_{+}}+\frac{1}{\Omega_{-}}\right).
\end{equation} 
This d.c. contribution to the electric current experiences the resonance behavior for the frequencies $\omega_J=\pm \omega_\parallel$ which should produce the steps in the current-voltage characteristics at $U=\pm \hbar \omega_\parallel /(2e)$ similar to the Shapiro steps arising under the effect of the microwave radiation.

\section{Inductance-induced magnetic anisotropy in the loop geometry}\label{Sec_Anisotropy}

The coupling between magnetic and plasma excitations in S/F$_1$/HM/F$_2$/S junctions gives rise to the peculiar renormalization of the magnetic anisotropy in ferromagnetic layer provided the $\psi$ junction is integrated into the closed superconducting loop (the geometry analogous to the rf SQUID). To demonstrate this we again consider the S/F$_1$/HM/F$_2$/S structure shown in Fig.~\ref{Fig1} where the spin quantization axis inside the half-metal coincides with the $z$ axis, the magnetization in the $F_1$ layer is directed along the $x$ axis, while the F$_2$ layer reveals the easy $xy$-plane magnetic anisotropy making the presence of the magnetization component $M_z$ energetically unfavorable. When such junction is embedded into the superconducting loop of the inductance $L$ the system Gibbs free energy takes the form \cite{}
\begin{equation}\label{FE2}
F=E_J\left[1+\cos\left(\varphi-\psi\right)+\frac{\varphi^2}{2\lambda}\right]+\frac{K_\perp V}{2}\left(\frac{M_z}{M_0}\right)^2,
\end{equation}
where $\lambda=2\pi cLI_c/\Phi_0$ is the dimensionless loop inductance, $K_\perp>0$ is the magnetic anisotropy constant, ${\bf M}$ is the magnetization vector in the F$_2$ layer.

Because of the finite loop inductance the system has only one ground state which corresponds to $M_z=M_y=0$, $M_x=-M_0$ ($\psi=\pi$) and $\varphi=0$. This is equivalent to the effective renormalization of the magnetic anisotropy: in the loop geometry there appears an additional weak easy-axis anisotropy along the $x$ axis.

To describe the dynamics of ${\bf M}$ out of equilibrium it is convenient to introduce the spherical coordinates in a way that $\theta$ is the angle between ${\bf M}$ and the $xy$-plane while $\psi$ is the polar angle in the $xy$-plane: $M_z=M_0\sin\theta$, $M_x=M_0\cos\theta\cos\psi$ and $M_y=M_0\cos\theta\sin\psi$. In the case of strong magnetic anisotropy one may assume $\theta\ll 1$.  The effective field (\ref{Heff}) takes the form
\begin{equation}\label{Heff_3}
{\bf H}_{eff}=-\frac{K_\perp M_z}{M_0^2}\hat{\bf z}_0-\frac{E_J}{V}\sin\left(\varphi-\psi\right)\frac{\partial \psi}{\partial {\bf M}},
\end{equation} 
where
\begin{equation}\label{gradPsi}
\frac{\partial \psi}{\partial {\bf M}}=\frac{M_x{\bf y}_0-M_y{\bf x}_0}{M_x^2+M_y^2}\approx\frac{1}{M_0^2}\left(M_x{\bf y}_0-M_y{\bf x}_0\right).
\end{equation} 
Then the Landau-Lifshitz-Gilbert equation (\ref{LLG}) can be written in the form of two coupled equations for the angles $\theta$ and $\psi$: 
\begin{equation}\label{thetapsi_eq}
\frac{\partial \theta}{\partial t}=\frac{\gamma E_J}{VM_0}\sin\left(\varphi-\psi\right),~~~\frac{\partial \psi}{\partial t}=-\frac{\gamma K_\perp}{M_0}\theta.
\end{equation}
At the same time, the dynamics of the Josephson phase is described the RSJ model equation accounting the finite loop inductance:
\begin{equation}\label{RSJ_L}
\frac{\partial^2\varphi}{\partial t^2}+\frac{\omega_p}{Q}\frac{\partial\varphi}{\partial t}+\omega_{LC}^2\varphi-\omega_p^2\sin\left(\varphi-\psi\right)=0,
\end{equation}
where $\omega_{LC}=\left(LC\right)^{-1/2}$ is the resonance frequency of the $LC$-circuit.

Considering the $e^{i\omega t}$ processes in the absence of damping ($Q\to\infty$) and linearizing Eqs.~(\ref{thetapsi_eq})-(\ref{RSJ_L}) near the ground state assuming the values $\varphi$, $\theta$ and $\delta\psi=\psi-\pi$ to be small we find the following characteristic equation which determines the resonance frequencies of the loop with embedded $\psi$ junction:
\begin{equation}\label{disp_perp}
\omega^4-\omega^2\left(\omega_{LC}^2+\omega_p^2+\Omega_\perp^2\right)+\Omega_\perp^2\omega_{LC}^2=0,
\end{equation}
where $\Omega_\perp^2=\gamma^2 E_J K_\perp/(VM_0^2)$. The equation (\ref{disp_perp}) always have two real solutions:
\begin{equation}\label{disp_perp_sol}
\omega^2=\frac{\omega_{LC}^2+\omega_p^2+\Omega_\perp^2\pm\sqrt{\left(\omega_{LC}^2+\omega_p^2+\Omega_\perp^2\right)^2-4\Omega_\perp^2\omega_{LC}^2}}{2}.
\end{equation}

For the loops with large inductance (when the frequency $\omega_{LC}$ is smaller than all other frequencies) the small collective magnetic and plasma excitations are characterized by two resonance frequencies:
\begin{equation}\label{perp_omega1}
\omega_1^2\approx\Omega_\perp^2+\omega_p^2+\omega_{LC}^2\frac{\omega_p^2}{\Omega_\perp^2+\omega_p^2},~~~\omega_2^2\approx \omega_{LC}^2\frac{\Omega_\perp^2}{\Omega_\perp^2+\omega_p^2}.
\end{equation}
Thus, the $LC$-circuit resonance frequency of rf SQUID based on $\psi$ junctions becomes dependent not only on the geometry of the Josephson junction (affecting its capacity) but also on its current-phase relation.

\section{Manipulation of magnetic moment by Josephson current}\label{Sec_magn_control}

The coupling between the magnetization direction and the Josephson phase in $\psi$ junctions allows the manipulation of magnetic moment by external d.c. current flowing through the junction. To illustrate the mechanism of such magnetic moment manipulation let us again consider the S/F$_1$/HM/F$_2$/S junction shown in Fig.~\ref{Fig1}. For simplicity we consider the situation when the magnetization in the F$_2$ layer has strong $xy$ easy-plane anisotropy and additional weak easy axis $x$ anisotropy. This ensures that the critical current of the Josephson junction remains almost constant since ${\bf M}$ tends to remain in the $xy$ plane but, at the same time, there are two equilibrium magnetization directions, namely, parallel and anti-parallel to the $x$ axis.   

In the presence of the external electric current $I$ flowing through the junction the system energy reads \cite{Likharev}
\begin{equation}\label{F_pa}
E=E_J\left[1+\cos\left(\varphi-\psi\right)-\frac{\varphi I}{I_c}\right]+\frac{V\left(K_\perp M_z^2-K M_x^2\right)}{2M_0^2},
\end{equation} 
where $K_\perp>0$, $K>0$, and we assume that $K_\perp\gg K,E_J/V$ which guarantees that the dynamics of the magnetic moment in the F$_2$ layer occurs in the vicinity of the $xy$ plane so that the critical current $I_c\approx const$. 

\begin{figure}[t!]
\includegraphics[width=0.48\textwidth]{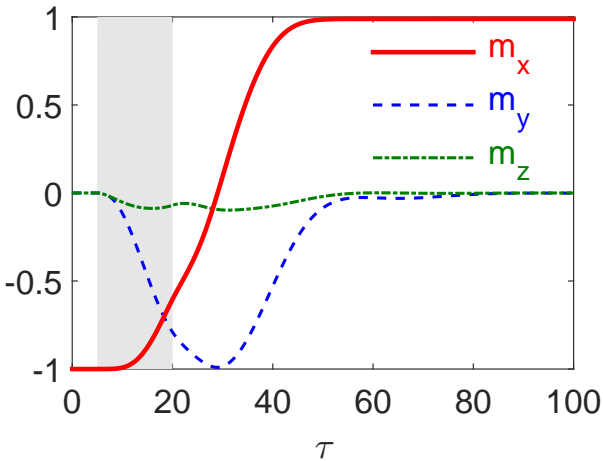}
\caption{The reversal of the magnetization direction under the effect of the current pulse. For calculations we took $\alpha=0.1$, $\kappa=0.02$, $\beta_\perp=0.02$, and $w=5$. The grey region shows the time interval when the external current $I=2I_c$ was applied. After $\tau=20$ the current was switched off.}  \label{Fig2}
\end{figure}

The dynamics of the magnetization ${\bf M}$ is described by Eq.~(\ref{LLG}) with the following effective field ${\bf H}_{eff}$:
\begin{equation}\label{Heff_manip}
{\bf H}_{eff}=\frac{KM_x}{M_0^2}{\bf x}_0-\frac{K_\perp M_z}{M_0^2}{\bf z}_0-\frac{E_J}{V}\sin(\varphi-\psi)\frac{M_x{\bf y}_0-M_y{\bf x}_0}{M_x^2+M_y^2}.
\end{equation}
In the case $K_\perp\gg K,E_J/V$ one may neglect the contributions $\propto (M_z/M_0)^2$ and put $M_x^2+M_y^2\approx M_0^2$. Then introducing the unit vector ${\bf m}={\bf M}/M_0$ and dimensionless time variable $\tau=(\gamma K_\perp t/M_0)(1+\alpha^2)^{-1}$ we obtain the following set of equations:
\begin{equation*}\label{Eq_mx}
\dot m_x=m_ym_z+\alpha\kappa m_x m_y^2-\beta_\perp\sin(\varphi-\psi)\left(m_xm_z-\alpha m_y\right),
\end{equation*}
\begin{equation*}\label{Eq_my}
\dot m_y=-(1+\kappa)m_xm_z-\alpha\kappa m_x^2 m_y-\beta_\perp\sin(\varphi-\psi)\left(m_ym_z+\alpha m_x\right),
\end{equation*}
\begin{equation}\label{Eq_mz}
\dot m_z=\kappa m_xm_y-\alpha(1+\kappa) m_x^2 m_z-\alpha m_y^2 m_z+\beta_\perp\sin(\varphi-\psi),
\end{equation}
where $\beta_\perp=E_J/(K_\perp V)\ll 1$, $\kappa=K/K_\perp\ll 1$, and the dot symbol denotes the derivative with respect to $\tau$.

In the limit when the Josephson junction has low capacity $C$ the relation between the transport current $I$ flowing through the junction and the Josephson phase $\varphi$ takes the form\cite{Likharev}
\begin{equation}\label{I_def}
\frac{I}{I_c}=\frac{1}{\omega_p Q}\frac{\partial \varphi}{\partial t}-\sin(\varphi-\psi).
\end{equation}
Introducing the dimensionless current $i=I/I_c$ and considering the dynamics in dimensionless time $\tau$ one may rewrite Eq.~(\ref{I_def}) in the form
\begin{equation}\label{I_mod}
i=w\dot\varphi-\sin(\varphi-\psi),
\end{equation}
where 
\begin{equation}\label{w_def}
w=\frac{1}{1+\alpha^2}\frac{\gamma K_\perp}{M_0}\frac{1}{\omega_pQ}.
\end{equation}
The above equations (\ref{Eq_mz}) and (\ref{I_mod}) determine the collective dynamics of the $\psi$ junction under the effect of external current. 

Interestingly, the application of the current pulse to the $\psi$ junction may result in the reversal of the magnetization direction. Fig.~\ref{Fig2} illustrates such reversal process. In the initial equilibrium state the magnetization is anti-parallel to the $x$ axis so that $m_x=-1$. Then at $\tau=5$ the transport current flowing through the junction becomes switched from zero to $I=2I_c$, remains constant till $\tau=20$ and then is switched off. The subsequent relaxation dynamics ends up in the new magnetic state with $m_x=1$. For typical S/F/S structures with $\gamma K_\perp/M_0\sim 10^{11}~{\rm sec}^{-1}$ the timescale of the magnetic moment switching is of the order of $10^{-10}~{\rm sec}$. Note that application of the second pulse will return the magnetic moment back to the state with $m_x=-1$. The described mechanism of magnetic switching driven by the current pulse can be used, e.g., provides a possibility to perform the precise control of magnetic state in the devices of superconducting spintronics.

\section{Conclusion}\label{Sec_conclusion}

To sum up, we predict the hybridization of magnetic moment and Josephson phase oscillations in S/F/HM/F/S junctions. Such coupling between two types of excitations does not require the propagation of the wave along the junction and occurs even in the long wavelength limit in contrast to the conventional S/F/S junctions. This provides the possibility to observe the collective excitations by the shift in the ferromagnetic resonance (FMR) frequency. In typical S/F/S junctions $\omega_\parallel\sim 10^{11}~{\rm s}^{-1}$ and $\omega_p\sim 10^{10}{~\rm s}^{-1}$ so that $\omega_p\lesssim\omega_\parallel$. At the same time, the parameter $\beta_\parallel=E_J/(K_\parallel V)$ may vary in the broad range from small values typical for the junctions with large distance between the superconducting electrodes (which damps both $E_J$ and $V^{-1}$) up to $\beta_\parallel\sim 100$ for the permalloy weak links \cite{Konschelle}. Thus, the increase of the FMR frequency due to the coupling with Josephson excitations can be comparable with $\omega_\parallel$. Note, that the very recent measurements showed the anomalously large FMR frequency shifts in S/F/S junctions \cite{Golovchanskiy, Li, Blamire_1, Blamire_2}. We hope that application of the similar experimental techniques to the S/F/HM/F/S junctions with alternating F/HM/F weak link thickness would provide the direct evidence of the predicted collective magnetic end plasma excitations in $\psi$ junctions.
 
Also, we find that integration of the S/F/HM/F/S $\psi$ junction into the superconducting loop can substantially change the effective magnetic anisotropy in one of the F layers. For example, in addition to the initial easy plane anisotropy the F layer can acquire the weak easy axis anisotropy which favors the specific orientation of the magnetic moment in the easy plane. In this case the coupling between magnetic and plasma oscillations also results in the shift of the resonance frequency of the $LC$-circuit which becomes dependent not only on the geometrical parameters of the Josephson junction but also on its current-phase relation. 

Finally, we have demonstrated the possibility to reverse the magnetization direction in one of the F layers of the S/F/HM/F/S structure by applying the pulse of the d.c. current. Interestingly, such reversal is accompanied by the switching between the $0$ and $\pi$ state of the Josephson junction since the ground-state phase is determined by the magnetic moment orientation. This provides a promising mechanisms for the controllable manipulation of magnetic moment and current-phase relation in the Josephson devices of superconducting spintronics. 

Note that similar phenomena can arise also in the Josephson S/F/F$^\prime$/F/S systems where the exchange field in the central F$^\prime$ layer is slightly smaller than the Fermi energy. Such structures are relatively easy to fabricate as compared to the half-metal based sandwiches. However, in this case the dependence of the spontaneous Josephson phase $\psi$ on the angle between the magnetic moments in the two F layers becomes non-linear and can even reveal hysteresis behavior which should affect the resonance frequencies of the collective excitations as well as the specific regimes of the magnetic moment switching. Thus, it would be interesting to analyze the collective magnetic and plasma excitations in S/F/F$^\prime$/F/S structures.

\section*{Acknowledgements}

The work of S.V.M. was supported by Center of Excellence ``Center of Photonics'' funded by The Ministry of Science and Higher Education of the Russian Federation (contract No. 075-15-2020-906). The work of A.I.B. was supported by the French ANR OPTOFLUXONICS and EU COST CA16218 Nanocohybri.

\end{document}